\documentclass{article}

\usepackage{arxiv}

\usepackage[utf8]{inputenc} 
\usepackage[T1]{fontenc}    
\usepackage{hyperref}       
\usepackage{url}            
\usepackage{booktabs}       
\usepackage{amsfonts}       
\usepackage{nicefrac}       
\usepackage{microtype}      
\usepackage{lipsum}

\usepackage{amsfonts}
\usepackage{amsmath}
\usepackage{subcaption}
\usepackage{graphicx}
\usepackage{amssymb}
\usepackage[inkscapelatex=false]{svg}
\usepackage{mathtools}
\usepackage{xcolor}
\usepackage{newtxtext,newtxmath}
\usepackage{lipsum}
\usepackage{float}
\usepackage{algorithm}
\usepackage{algpseudocode}
\usepackage{stmaryrd}
\usepackage{algorithm}
\usepackage{algpseudocode}

\newcommand{\R}{{\mathbb{R}}}

\newcommand{\N}{{\mathbb{N}}}
\newcommand{\No}{{\mathcal{N}}}

\newcommand{\Io}{{\mathcal{I}}}

\newcommand{\ie}{{\it i.e.}}

\newcommand{\X}{{\mathbf{X}}}

\newcommand{\W}{{\mathbf{W}}}

\newtheorem{theorem}{Theorem}[section]
\newtheorem{assumption}{Assumption}

\newtheorem{definition}[theorem]{Definition}
\newtheorem{lemma}[theorem]{Lemma}
\newtheorem{remark}[theorem]{Remark}
\newenvironment{proof}{\paragraph{Proof:}}{\hfill$\square$}
\newtheorem{problem}[theorem]{Problem}
\usepackage{graphicx} 

\title{Scalable Formal Verification of Incremental Stability in Large-Scale Systems Using Graph Neural Networks}

\author{
 Ahan Basu \\
  Centre for Cyber-Physical Systems\\
  Indian Institute of Science, Bengaluru, India\\
  \texttt{ahanbasu@iisc.ac.in} \\
   \And
 Mahathi Anand \\
  Chair of Robotics and System Intelligence \\ Munich Institute of Robotics and Machine Intelligence \\ Technical University of Munich, Germany\\
  \texttt{mahathi.anand@tum.de} \\
  \And
 Pushpak Jagtap \\
  Centre for Cyber-Physical Systems\\
  Indian Institute of Science, Bengaluru, India\\
  \texttt{pushpak@iisc.ac.in} \\
}

\begin{document}

\maketitle

\begin{abstract}
    This work proposes a novel distributed framework for verifying the incremental stability of large-scale systems with unknown dynamics and known interconnection structures using graph neural networks. Our proposed approach relies on the construction of local incremental Lyapunov functions for subsystems, which are then composed together to obtain a suitable Lyapunov function for the interconnected system. Graph neural networks are used to synthesize these functions in a data-driven fashion. The formal correctness guarantee is then obtained by leveraging Lipschitz bounds of the trained neural networks. Finally, the effectiveness of our approach is validated through two nonlinear case studies.
\end{abstract}

\section{Introduction}
Incremental stability is an important property studied often in the field of control theory as it provides guarantees on the convergence of all trajectories with respect to each other than with respect to a fixed equilibrium point. Thus, it finds itself useful in several applications such as cyclic feedback system synchronization \cite{cyclic_feedback}, symbolic model development \cite{symbolic1, jagtap2020symbolic, traffic_model}, and complex network analysis \cite{synchComplex}. In the last few decades, researchers have developed several tools to analyse incremental stability, such as contraction theory \cite{Contraction}, convergent dynamics \cite{Conv_dyn} and incremental Lyapunov functions \cite{angeli2002lyapunov, DT-ISS}. These tools have been widely extended to analyse incremental stability properties in several classes of systems, such as stochastic systems \cite{jagtap2017backstepping}, hybrid systems \cite{biemond2018incremental}, switched interconnected systems \cite{dey2025incremental}, and time-delayed systems \cite{chaillet2013razumikhin}.  However, all of these methods rely on the assumption of complete knowledge of the system dynamics.

To deal with systems with unknown dynamics, learning-based approaches have gained significant attention in recent years. For example, deep neural networks can be utilized for the synthesis of Lyapunov functions ~\cite{Lyapunov_nn, formal_nn_Lyapunov}, omitting the necessity of precise mathematical models. However, providing formal guarantees for the learned Lyapunov functions has been a challenge since they are trained only over a finite samples collected from the state-space. To assert such guarantees, one usually trains the neural networks with \textit{sufficiently} many data samples and constructs sampling-based optimization problems, as in \cite{DD-Stability, basu2025formally}. However, such an approach is not suitable for large-scale interconnected systems as deep neural networks offer a centralized architecture to represent Lyapunov functions. Due to the exponential complexity associated with sampling of data, verification of incremental stability properties for large-scale systems would result in computational intractability. Furthermore, deep neural networks do not take advantage of the modular structure offered by interconnected systems, resulting in unnecessarily large communication overhead. A recent approach \cite{zhang2023compositional} has leveraged weight-sharing techniques for feedforward networks to compute local Lyapunov functions and uses the small-gain compositionality condition to verify ISS for the interconnected system. However, this approach can only be used for a class of affine control systems whose dynamics are known.

This article seeks to alleviate the aforementioned challenges by proposing a distributed framework to synthesize Lyapunov functions verifying incremental stability for large-scale discrete-time control systems in a scalable, data-driven fashion. To do so, we consider large-scale systems consisting of several unknown subsystems connected via some known graph-based topologies. First, we define the notion of \textit{so-called} local $\delta$-ISS Lyapunov functions that depend only on the information from neighbouring subsystems, thus avoiding the need for large communication overhead. Then, we propose a compositional framework to construct the $\delta$-ISS Lyapunov function for the large-scale {interconnected} system that verifies the {incremental stability} of the large-scale system via the local functions. We employ graph neural networks (GNNs) \cite{scarselli2008graph} to represent the local Lyapunov functions, which are also characterized by the same interconnection topology as the large-scale systems. As a result, unlike deep neural network approaches, graph neural networks inherently offer a distributed architecture, making them useful for the synthesis of $\delta$-ISS Lyapunov functions in a scalable manner. Furthermore, we provide a data-driven formal verification scheme to validate the correctness of the trained $\delta$-ISS Lyapunov functions by constructing a sampling-based verification technique that relies on Lipschitz bounds of the trained GNNs corresponding to the $\delta$-ISS local Lyapunov functions. Finally, we validate our proposed approach with two different case studies. 

While GNNs have been employed for a variety of distributed verification and control tasks, including adaptive control of multi-agent systems \cite{fallin2025lyapunov}, cooperative control of autonomous vehicles \cite{chen2021graph} and safety verification of unknown large-scale systems \cite{anand2024distributed}, we would like to note that, to the best of our knowledge, this is the first work to employ GNNs for the verification of \textit{incremental stability of large-scale systems} while providing formal guarantees. 

\section{Preliminaries and Problem Formulation}

\subsection{Notations}
The symbols $\N$, $ \N_0$, $ \R$, $\R^+$, and $\R_0^+$ denote the set of natural, nonnegative integers, real, positive reals, and nonnegative real numbers, respectively. For $a, b \in \N_0$ with $a \leq b$, the closed interval in $\N_0$ is denoted as $[a; b]$.
The vector space of real matrices with $ n $ rows and $ m $ columns is denoted by $\R^{n\times m} $. The column vector space with $n$ rows is represented by $ \R^{n}$.
The Euclidean norm is represented using $|\cdot |$.  
Given a function $\varphi: \N_0 \rightarrow \R^m$, its sup-norm is given by $\lVert \varphi \rVert = \sup\{|\varphi(k)| : k \in \N_0\}$.
Given a matrix $M\in\R^{n\times m}$, $M^\top$ represents the transpose of matrix $M$ {while $[M]_{ij}$ denotes the element corresponding to the $i$-th row and $j$-th column of the matrix $M$.}
A continuous function $\alpha: \R_0^+ \rightarrow \R_0^+$ is said to be class $\mathcal{K}$ if $\alpha(s)>0 $ for all $s>0$, strictly increasing and $\alpha(0)=0$. It is class $\mathcal{K}_\infty$ if it is class $\mathcal{K}$ and $\alpha(s)\rightarrow\infty$ as $s\rightarrow\infty$.
A continuous function $\beta: \R_0^+ \times \R_0^+ \rightarrow \R_0^+$ is said to be a class $\mathcal{KL}$ if $\beta(s,t)$ is a class $\mathcal{K}$ function with respect to $s$ for all $t$ and for fixed $s>0$, $\beta(s,t) \rightarrow 0 $ as $t\rightarrow \infty$. 

\subsection{Incremental Stability for Interconnected Systems}
In this article, we study the notion of interconnected systems composed of $N$ discrete-time nonlinear subsystems (dt-NHS) represented as $\Upsilon_i = (\hat{\X}, \hat{\W}, f_i)$, where the following equation governs the dynamics of individual subsystems:
\begin{equation}\label{eq:subsystem_dyn}
    \mathsf{x}_i(k+1) = f_i(\mathsf{x}_i(k), \{\mathsf{x}_j(k)\}_{j \in \No_i}, \mathsf{w}_i(k)), i \in \Io_N,
\end{equation}
where $\Io_N = \{1,\ldots,N\}, \mathsf{x}_i(k) \in \hat{\X} \subseteq \R^n$ is the state of $i$-th subsystem, $\mathsf{w}_i(k) \in \hat{\W} \subseteq \R^m$ is the external input to the $i$-th subsystem, $\No_i \in \Io_N$ denotes the set of {neighbouring subsystems that influence the $i^\text{th}$ subsystem,} and $f_i:\R^n \times \R^{(n \times {\No_i})} \times \R^m \rightarrow \R^n$ denotes the state transition function for the $i$-th subsystem {that follows
\begin{align}
    f_i(\mathsf{x}_i, \{\mathsf{x}_j\}_{j \in \No_i}, \mathsf{w}_i) = \bar{f}(\mathsf{x}_i, \mathsf{w}_i) + \sum_{j \in \mathcal{N}_i}h(\mathsf{x}_j, \mathsf{x}_i).
\end{align}}
Additionally, we consider each subsystem to be forward-invariant\footnote{If the system starts from initial condition $\mathsf{x}_i(0) \in \hat{\X}$, for any $k \in \N$, under any external input $\mathsf{w}_i(k) \in \hat{\W}$, the state of the system $\mathsf{x}_i(k) \in \hat{\X}$.} and locally Lipschitz. The state and input of the large-scale system are given by the stacked vectors $\mathsf{x}(k) = [\mathsf{x}_1(k); \ldots; \mathsf{x}_N(k)] \in \X:= \hat{\X}^N$ and $\mathsf{w}(k) = [\mathsf{w}_1(k); \ldots; \mathsf{w}_N(k)] \in \W:= \hat{\W}^N$, respectively.

Now, for the $i$-th subsystem, we denote the concatenated state vector and input vector as $\Tilde{\mathsf{x}}_i(k) \coloneq [\mathsf{x}_i(k), \{\mathsf{x}_j(k)\}_{j \in \No_i}]^\top$  $\in \Tilde{\X}_i:=\hat{\X}^{\No_i+1}$ and $\Tilde{\mathsf{w}}_i(k) \coloneq [\mathsf{w}_i(k), \{\mathsf{w}_j(k)\}_{j \in \No_i}]^\top  \in \Tilde{\W}_i:=\hat{\W}^{\No_i+1}$ respectively. Consequently, the subsystem dynamics is denoted as:
\begin{align}\label{eq:subsystem_dyn_2}
    \Tilde{\mathsf{x}}_i(k+1) = \Tilde{f}_i(\Tilde{\mathsf{x}}_i(k), \Tilde{\mathsf{w}}_i(k)), i \in \Io_N.
\end{align}
where $\Tilde{f}_i = f_i \times \prod_{j \in \No_i}f_j$. This results in the interconnected system given by
\begin{align}\label{eq:complete}
    \Upsilon: \mathsf{x}(k+1) = f(\mathsf{x}(k), \mathsf{w}(k)), \hspace{0.1em} \text{with} \hspace{0.3em} f:\X \times \W \rightarrow \X,
\end{align}
{with a homogenous interconnection topology induced by a} static, directed graph $\mathcal{G} = (\mathcal{V}, \mathcal{E})$ where $\mathcal{V} = \{v_1,\ldots,v_N\}$ are the vertices of $\mathcal{G}$ with $v_i, i \in \mathcal{I}_N$ representing the $i$-th subsystem and $\mathcal{E}:=\{(v_i,v_j)\}$ where $i$-th subsystem influences $j$-th subsystem. {We define the adjacency matrix $\mathsf{A}$ consistent with the interconnection topology such that $[\mathsf{A}]_{ii}=1$ for all $i \in \mathcal{I}_N$ while $[\mathsf{A}]_{ij}=1$ if $(v_i, v_j) \in \mathcal{E}, i \neq j$ and the remaining elements are zero.} As defined earlier, $\mathcal{N}_i$ is the set of neighbours of node $v_i$.

We denote the state of the {overall interconnected} system at time instance $k$ as $\mathsf{x}_{x,\mathsf{w}}(k)$ with the system starting from the initial state $x \in \X$ under the sequence of input $\mathsf{w} \in \W$. In the following, we define the notion of incremental stability for the interconnected system.

\begin{definition}\label{def:ISS_int}
    An interconnected dt-NHS $\Upsilon$ as in \eqref{eq:complete} is said to be incrementally input-to-state stable ($\delta$-ISS) if 
    \begin{equation}\label{eq:iss-system}
        |\mathsf{x}_{x,\mathsf{w}}(k)-\mathsf{x}_{\hat x,\hat{\mathsf{w}}}(k)| \leq \beta(|x-\hat x|,k) + \rho(\lVert \mathsf{w} - \hat{\mathsf{w}} \rVert),
    \end{equation}
    for any $x, \hat{x} \in \X, w, \hat{w} \in \W$ and some $\beta \in \mathcal{KL}, \rho \in \mathcal{K}_{\infty}$. Note that, under the condition $\mathsf{w} = \hat{\mathsf{w}} = 0$ and $\X = \R^n$, one can recover the definition of incremental global asymptotic stability, wherein any two trajectories converge to origin over time.
\end{definition}
We now utilize $\delta$-ISS Lyapunov functions to prove $\delta$-ISS properties via the following theorem. 
\begin{theorem}\label{th:cent_Lyap}
    The interconnected dt-NHS $\Upsilon$ as in \eqref{eq:complete} is said to be incrementally input-to-state stable ($\delta$-ISS) if there exists a function $V:\R^{N\times n} \times \R^{N \times n} \rightarrow \R$ of degree $\kappa \in \N^+$ with class $\mathcal{K}_\infty$ functions $\underline \alpha, \overline{\alpha}, \Tilde{\alpha}$ and a class $\mathcal{K}$ function $\sigma$ such that for all $x,\hat{x}\in \X$ and for all $w, \hat{w} \in \W$:
    \begin{subequations}\label{eq:ISS-Lf}
    \begin{align}
        \underline \alpha(|x-\hat{x}|) &\leq V(x,\hat{x}) \leq \overline{\alpha}(|x-\hat{x}|), \label{eq:Lyap_bound} \\
        V(f(x, w),f(\hat{x},\hat{w})) &- V(x,\hat{x}) \leq -\Tilde{\alpha}(|x - \hat{x}|) + \sigma(|w-\hat{w}|). \label{eq:Lyap_diff}
    \end{align}
    \end{subequations}
\end{theorem}
\begin{proof}
    The proof is similar to that of \cite[Theorem 1]{DT-ISS} and hence omitted here.
\end{proof}

\subsection{Problem Formulation}
This article deals with interconnected systems $\Upsilon$ with unknown dynamics $f$ and known topological structures. This is supported by the assumption stated below.

\begin{assumption}\label{assum:black-box}
    {For the interconnected system $\Upsilon$,} we assume that the interconnection topology $\mathcal{G}$ and its corresponding adjacency matrix $\mathsf{A}$ are known. Moreover, while the functions $f_i$, $i \in \mathcal{I}_N$ for the subsystems are unknown, we assume access to the black-box model, i.e., given state-input pair of the system $(x,w) \in \X \times \W$, one can determine the next state $f(x,w)$ of the interconnected system.
\end{assumption}

Now, we are ready to state the main problem of the paper. 
\begin{problem}
    Consider a large-scale {interconnected} system $\Upsilon$ as in \eqref{eq:complete} composed of $N$ subsystems interconnected via the graph-based topology $\mathcal{G}$ and governed by \eqref{eq:subsystem_dyn_2}, satisfying Assumption~\ref{assum:black-box}.
    The objective is to construct a valid $\delta$-ISS Lyapunov function (if it exists) for $\Upsilon$ to certify incremental input-to-state stability properties. 
\end{problem}

{Since we deal with} large-scale, unknown, interconnected systems, certain challenges prevent us from directly synthesizing the $\delta$-ISS Lyapunov certificate. First,  condition \eqref{eq:Lyap_diff} can not be directly leveraged due to its dependence on the system dynamics $f$, which is unknown. Secondly, due to limited communication capabilities, the subsystems may not have access to the states of all the other subsystems, but only the information of their neighboring subsystems. Therefore, it is of interest to obtain a distributed, data-driven framework for the construction of $\delta$-ISS Lyapunov functions ensuring the incremental ISS of the large-scale system.

\section{Distributed Construction of $\delta$-ISS Lyapunov Functions}\label{sec:constr}

This section focuses on constructing a $\delta$-ISS Lyapunov function to formally ensure the incremental stability of an interconnected system with unknown dynamics. In particular, we formulate a distributed approach for the construction of $\delta$-ISS Lyapunov functions by designing \emph{local $\delta$-ISS functions} for each subsystem and composing them together to prove $\delta$-ISS properties of the interconnected system. 

\begin{definition}\label{def:l-ISS-LF}
    Consider a large-scale interconnected system as in \eqref{eq:complete} that is composed of $N$ subsystems as in \eqref{eq:subsystem_dyn}, under the interconnection topology $\mathcal{G} = (\mathcal{V}, \mathcal{E})$. Then, for a subsystem $\Upsilon_i$ as in \eqref{eq:subsystem_dyn} corresponding to a node $v_i \in \mathcal{V}, i \in \Io_N$, a function $V_i: \R^n \times \R^n \rightarrow \R_0^+$ is a local $\delta$-ISS Lyapunov function of degree $\kappa \in \N$ if the following conditions hold:
    \begin{subequations}\label{eq:l-ISS-LF}
    \begin{align}
        &\forall \Tilde{x}_i, \hat{\Tilde{x}}_i \in \Tilde{\X}_i: \underline \alpha_i |\Tilde{x}_i - \hat{\Tilde{x}}_i|^\kappa \leq V_i(\Tilde{x}_i,\hat{\Tilde{x}}_i) \leq \Bar{\alpha}_i |\Tilde{x}_i - \hat{\Tilde{x}}_i|^\kappa  \label{eq:bounds} \\
        &\forall \Tilde{x}_i, \hat{\Tilde{x}}_i \in \Tilde{\X}_i, \forall \Tilde{w}_i, \hat{\Tilde{w}}_i \in \Tilde{\W}_i: V_i(\Tilde{f}_i(\Tilde{x}_i, \Tilde{w}_i), \Tilde{f}_i(\hat{\Tilde{x}}_i,\hat{\Tilde{w}}_i)) - V_i(\Tilde{x}_i,\hat{\Tilde{x}}_i) \leq -\Tilde{\alpha}_i|\Tilde{x}_i - \hat{\Tilde{x}}_i|^\kappa + \sigma_i|\Tilde{w}_i-\hat{\Tilde{w}}_i|^\kappa \label{eq:diff}
    \end{align}
    \end{subequations}
    for some $\underline \alpha_i, \Bar{\alpha}_i, \Tilde{\alpha}_i \in \R^+, \sigma_i \in \R_0^+$.
\end{definition}

\begin{remark}
    {Without loss of generality, the comparison bounds associated with $V_i$, namely the lower and upper bounds in \eqref{eq:bounds} and the upper bound in \eqref{eq:diff}, are polynomials of degree $\kappa$. The specific value of $\kappa$ is chosen during training to determine an appropriate form of $V_i$.}
\end{remark}

We are interested in constructing local $\delta$-ISS Lyapunov functions $V_i$ to locally satisfy condition \eqref{eq:l-ISS-LF} over some compact state-space $\X$ which is assumed to be forward invariant. Then we show that the Lyapunov function $V$ for the interconnected system can be constructed using the local $\delta$-ISS Lyapunov functions. To do so, we first propose the following lemma:

\begin{lemma}\label{lem:classk}
    The sum of {$p (\geq 1)$} class $\mathcal{K}_{\infty}$ functions will be a class $\mathcal{K}_{\infty}$ function.
\end{lemma}

\begin{proof}
    {The result for $p=1$ is trivial. For $p=2$, consider }$\mathcal{K}_{\infty}$ functions $\xi_1(x)$ and $\xi_2(x)$. Clearly, $\xi_1(0) = \xi_2(0) = 0, \frac{d \xi_1(x)}{dx}>0,\frac{d \xi_2(x)}{dx}>0$ and as $x \rightarrow \infty, \xi_1(x), \xi_2(x) \rightarrow \infty$. 
    Let, $\xi(x) = \xi_1(x) + \xi_2(x)$. Clearly, $\xi(0) = \xi_1(0) + \xi_2(0) = 0.$ $\frac{d \xi(x)}{dx} = \frac{d \xi_1(x)}{dx} + \frac{d \xi_2(x)}{dx} >0$ and as $x \rightarrow \infty, \xi(x) \rightarrow \infty$. hence, $\xi(x)$ is also a class $\mathcal{K}_{\infty}$ function. {By applying induction, one can extend the proof for the sum of $p > 2$ class $\mathcal{K}_{\infty}$ functions.}
\end{proof}

Now we are ready to present the main result of this subsection.
\begin{theorem}\label{th:composition}
    Consider a large-scale system governed by \eqref{eq:complete} composed of $N$ homogeneous subsystems as in \eqref{eq:subsystem_dyn_2}, under the interconnection topology $\mathcal{G}$. Suppose for all $i \in \Io_N$, there exists a local $\delta$-ISS Lyapunov function satisfying the conditions of Definition \ref{def:l-ISS-LF}. Then, $V(x) = \sum_{i=1}^N V_i(\Tilde{x}_i, \hat{\Tilde{x}}_i)$ is a $\delta$-ISS Lyapunov function for the large-scale system.
\end{theorem}

\begin{proof}
    Suppose that there exists a local $\delta$-ISS Lyapunov function for each subsystem of the large-scale system satisfying {conditions~\eqref{eq:l-ISS-LF}}. Then, we {construct a} $\delta$-ISS Lyapunov function for the {interconnected system by considering $V(x, \hat x) = \sum_{i=1}^N V_i(\Tilde{x}_i, \hat {\Tilde{x}}_i)$ and obtaining the following bounds:}
    \begin{align*}
    \text{(a)} &\ V(x, \hat{x})\! =\!  \sum_{i=1}^N V_i(\Tilde{x}_i, \hat{\Tilde{x}}_i)\!  \geq \! \sum_{i=1}^N  \underline \alpha_i |\Tilde{x}_i - \hat{\Tilde{x}}_i|^\kappa := \underline \alpha (|x - \hat{x}|) \\
    \text{(b)} &\ V(x, \hat{x}) \! = \! \sum_{i=1}^N V_i(\Tilde{x}_i, \hat{\Tilde{x}}_i) \! \leq \! \sum_{i=1}^N  \Bar{ \alpha}_i |\Tilde{x}_i - \hat{\Tilde{x}}_i|^\kappa := \overline{ \alpha} (|x - \hat{x}|) \\
    \text{(c)} &\ V(f(x, w),f(\hat{x},\hat{w})) - V(x,\hat{x}) = \sum_{i=1}^N \big(V_i(\Tilde{f}_i(\Tilde{x}_i, \Tilde{w}_i), \Tilde{f}_i(\hat{\Tilde{x}}_i, \hat{\Tilde{w}}_i)) - V_i(\Tilde{x}_i,\hat{\Tilde{x}}_i) \big) \\
    & \leq \sum_{i=1}^N -\Tilde{\alpha}_i|\Tilde{x}_i - \hat{\Tilde{x}}_i|^\kappa + \sum_{i=1}^N \sigma_i|\Tilde{w}_i-\hat{\Tilde{w}}_i|^\kappa := -\Tilde{\alpha}(|x - \hat{x}|) + \Tilde{\sigma}(|w-\hat{w}|). 
    \end{align*}
    Clearly, the function $V(x, \hat{x}):= \sum_{i=1}^N V_i(\Tilde{x}_i, \hat{\Tilde{x}}_i)$ is a $\delta$-ISS Lyapunov function for the large-scale system as in \eqref{eq:complete} and guarantees the incremental input-to-state stability of the system.
\end{proof}

\section{GNN-based Lyapunov Functions for Unknown Interconnected Systems}\label{sec:NN}

To synthesize $\delta$-ISS Lyapunov functions in a distributed manner via Theorem~\ref{th:composition}, one still needs knowledge of the system dynamics, as the local $\delta$-ISS Lyapunov functions depend on the subsystem dynamics $f_i$ as in~\eqref{eq:subsystem_dyn}-\eqref{eq:subsystem_dyn_2}. Therefore, to deal with unknown dynamics, we propose a data-driven, distributed computational framework for the construction of Lyapunov functions. To this end, we utilize graph neural network-based architectures for representing local $\delta$-ISS Lyapunov functions, as they inherently provide a distributed architecture that is aligned with the interconnection topology of the large-scale system.

First, we discuss the GNN architecture utilized for parameterizing Lyapunov functions, following which a training scheme via carefully crafted loss functions is obtained. Finally, we provide a formal verification approach to validate the correctness guarantees over the trained Lyapunov functions by leveraging Lipschitz continuity properties of the GNNs.

\subsection{Graph Neural Network Architecture}\label{subsec:GNN_archi}
Consider the graph $\mathcal{G}$ of the interconnected system \eqref{eq:complete} and its corresponding adjacency matrix $\mathsf{A}$. To precisely describe the input-output relationships in the graph neural network architecture, we first reformulate the state $x$ of the large-scale system as an {$N \times n$ matrix $\mathbf{x} = [x_1, \ldots, x_N]^\top$}, where the $i$-th row of the matrix represents the state of the  $i$-th subsystem. Now using the adjacency matrix $\mathsf{A}$ the state information from $\mathbf{x}$ can be shifted across the graph as:
\begin{align}
    [\mathsf{A}\mathbf{x}]_{ik} = \sum_{j=1}^N[\mathsf{A}]_{ij}[\mathsf{x}]_{jk} = [\mathsf{A}]_{ii}[\mathsf{x}]_{ik} + \sum_{j \in \mathcal{N}_i}[\mathsf{A}]_{ij}[\mathsf{x}]_{jk}.
\end{align}
From the above equation, it can be seen that for each node $i$ of the graph representing the subsystem $i$, the state information from that subsystem as well as its neighboring nodes are aggregated. 

Utilizing the {above} operation to gather information across neighbouring nodes, the GNN corresponding to the $\delta$-ISS Lyapunov function is constructed by means of graph filter \cite{Gama2019GNN}, feeding it to a pointwise nonlinear activation function $\phi(\cdot)$ (for e.g., ReLU) and cascading $\Bar{l}_b$ such layers together. The output of $\Bar{l}_b$-th layer at every node is then fed to a standard fully connected neural network with $l_b$ layers and $h_b^j$ neurons in each layer $j \in \{\Bar{l}_b, \ldots, \bar{l}_b + l_b -1\}$. 

Therefore, we summarize the GNN architecture as follows:
$\begin{cases}
    \mathsf{x}^0 = \mathsf{x} \\
    \psi(\mathsf{x}^l, \mathsf{A}) = \mathsf{x}^l\mathbf{H}_l^0 + \mathsf{A}\mathsf{x}^l\mathbf{H}_l^1,\quad l \in [0; \bar{l}_b-1], \\
    \mathsf{x}^{l+1} = \phi(\psi(\mathsf{x}^l, \mathsf{A})), \quad l \in [0; \bar{l}_b-1],\\
    [\mathsf{x}^{l+1}]_i = \phi(\mathbf{W}^l[\mathsf{x}^l]_i + b^l), \quad l \in [\bar{l}_b; \bar{l}_b + l_b-1],\\
    \hspace{4.1cm} i \in \mathcal{I}_N, \\
    [\mathsf{x}^{\bar{l}_b+l_b+1}]_i = \mathbf{W}^{\bar{l}_b+l_b}[\mathsf{x}^{\bar{l}_b+l_b}]_i + b^{\bar{l}_b+l_b}, \quad i \in \mathcal{I}_N.
\end{cases}$

The filter coefficient matrices that characterize the graph filter is denoted by $\mathbf{H}_l^k, \ k \in \{0,1\}$ whose dimensions are $z_{l} \times z_{l+1}$ for every layer $l \in \{0, \ldots, \bar{l}_b-1\}$ while the input and output dimension of the graph filtering layer is denoted by $N \times z_{l}$ with $z_0 = n$, while the parameters $\mathbf{W}^l$ and $b^l$ are the weight and bias coefficients of the fully connected network. Therefore, for a subsystem corresponding to node $v_i, \ i \in \mathcal{I}_N$, the local $\delta$-ISS Lyapunov function $V_i(\Tilde{x}_i) = [\mathsf{x}^{\bar{l}_b+l_b+1}]_i$ which is the output of the {$i^\text{th}$ node of} GNN.

\subsection{Training Procedure} \label{subsec:training}
{Consider} $\delta$-ISS Lyapunov functions $V_i^{\gamma}$ for all $i \in \mathcal{I}_N$
{represented by a GNN such that} $V^{\gamma}$ is parametrized by the training parameters $\gamma = [\mathbf{H}, \mathbf{W}, b]$. Then, it is evident from Theorem \ref{th:composition} that the $\delta$-ISS Lyapunov function is of the form $V^\gamma(x, \hat{x}) = \sum_{i=1}^N V_i^{\gamma}(\Tilde{x}_i, \hat{\Tilde{x}}_i)$. Now, to train the neural network, we first propose to characterize the conditions \eqref{eq:l-ISS-LF} as loss functions so that they are minimized over the training datasets. 

{For the large-scale control system, consider a finite set of training data points obtained from the system by randomly collecting $M$ points from the augmented space $D = (\X \times \X \times \W \times \W)$, which is denoted by $\mathcal{D}:=\{y_j = (x_j, \hat{x}_j, w_j, \hat{w}_j) \mid j \in [1;M], y_j \in D\}$}. Then, we construct the sub-loss functions:
\begin{subequations}\label{eq:sub_loss_func}
\begin{align} 
    L_1(\gamma) &= \hspace{-0.2cm}\sum_{y \in \mathcal{D}} \sum_{i=1}^N \text{ReLU}(-V_i^{\gamma}(\Tilde{x}_{is}, \hat{\Tilde{x}}_{is}) + \underline \alpha_i |\Tilde{x}_{is} - \hat{\Tilde{x}}_{is}|^\kappa - \lambda_i), \\
    L_2(\gamma) &= \sum_{y \in \mathcal{D}} \sum_{i=1}^N \text{ReLU}(V_i^{\gamma}(\Tilde{x}_{is}, \hat{\Tilde{x}}_{is}) - \overline \alpha_i |\Tilde{x}_{is} - \hat{\Tilde{x}}_{is}|^\kappa - \lambda_i), \\
    L_3(\gamma) &= \sum_{y \in \mathcal{D}} \sum_{i=1}^N \text{ReLU}(V_i^{\gamma}(\Tilde{f}_i(\Tilde{x}_{is}, \Tilde{w}_{is}), \Tilde{f}_i(\hat{\Tilde{x}}_{is}, \hat{\Tilde{w}}_{is})) - V_i^\gamma(\Tilde{x}_{is}, \hat{\Tilde{x}}_{is}) +\Tilde{\alpha}_i|\Tilde{x}_{is} - \hat{\Tilde{x}}_{is}|^\kappa - \sigma_i|\Tilde{w}_{is}-\hat{\Tilde{w}}_{is}|^\kappa - \lambda_i),
\end{align}
\end{subequations}
where $\lambda_i < 0, \ i \in \mathcal{I}_N$ are the generalization parameters used to ensure the local $\delta$-ISS Lyapunov functions can be generalized to values outside the training dataset. We minimize the loss function:
\begin{align}\label{eq:loss}
    L(\gamma) = c_1L_1(\gamma) + c_2L_2(\gamma) + c_3L_3(\gamma),
\end{align}
where $c_1, c_2, c_3 \in \R^+$ are weights corresponding to the sub-loss functions. One can readily observe that if the training converges with a final loss $L=0$, then the local $\delta$-ISS Lyapunov conditions \eqref{eq:l-ISS-LF} have been satisfied over finitely many data points, implying the $\delta$-ISS Lyapunov function
{ $ V(x,\hat x) = \sum_{i=1}^n V^\gamma(\Tilde{x}_i, \hat {\Tilde{x}}_i)$ is a suitable candidate} for the verification of the ISS properties of the large-scale interconnected system.

{Training via graph neural networks proposed above possesses several advantages. First, GNNs possess transferability properties to larger graphs. so long as the interconnection topology remains similar (e.g. properties such as the number of neighbors and symmetry are preserved\cite{Ruiz2020graphon}). This enables training the GNNs for a smaller interconnected structure with fewer data samples, and then fine-tuning the network for a larger interconnected system without any requirement to retrain from scratch, aiding scalability in the training procedure. Furthermore, the permutation equivariance property of the GNN suggests that the training is stable under node reordering,~\cite{Gama2022Distributed}, enabling generalization of the training framework to unknown yet similar graphs.}

{However, the candidate function obtained via the aforementioned training function is not formally guaranteed to be correct since} the GNN is trained over {only finitely many} data points, and not on the complete state-space $\X$. In the next subsection, we present the technique to formally verify the correctness of the candidate $\delta$-ISS Lyapunov function.

\subsection{Formal Correctness Guarantees}\label{subsec:guarantee}
Since the trained network $V^{\gamma}(x, \hat{x})$ satisfies the conditions over {only finitely many} random samples collected from the state-space, one needs to formally verify whether the local $\delta$-ISS Lyapunov functions satisfy condition \eqref{eq:l-ISS-LF} and subsequently the \eqref{eq:ISS-Lf} over the entire state space.

{To do so, we leverage the Lipschitz continuity of the learned $\delta$-ISS Lyapunov function as well as the system dynamics and propose a sampling-based verification approach for verifying the validity of the candidate function.} 
Specifically, for the $i$-th subsystem of the large-scale system \eqref{eq:complete}, we approximate the state space $\Tilde{\X}_i$ and input space $\Tilde{\W}_i$ using a finite number of samples. Specifically, we draw $R$ samples $\Tilde{x}_{ir}$ from $\Tilde{\X}_i$, indexed by $r \in [1; R]$. 
Around each sample $\Tilde{x}_{ir}$, we construct a ball $B_{\varepsilon_{ix}}(\Tilde{x}_{ir})$ of radius $\varepsilon_{ix}$ such that for every $\Tilde{x}_i \in \Tilde{\X}_i$ there exists a sample $\Tilde{x}_{ir}$ satisfying $|\Tilde{x}_i - \Tilde{x}_{ir}| \leq \varepsilon_{ix}$. This guarantees that $\bigcup_{r=1}^R B_{\varepsilon_{ix}}(\Tilde{x}_{ir}) \supset \Tilde{\X}_i$. An analogous procedure is used for the input space $\Tilde{\W}_i$, where we collect $P$ samples $\Tilde{w}_{ip}$ and form balls of radius $\varepsilon_{iu}$ around them. Collecting the data points obtained upon sampling the state-space $\Tilde{\X}_i$ and input space $\Tilde{\W}_i$, {we form the verification datasets as}:
\begin{align}\label{set:SCP}
\mathcal{X}_i \!=\! \{\Tilde{x}_{ir} | \!\bigcup_{r=1}^{R} \! B_{\varepsilon_{ix}}(\Tilde{x}_{ir}) \!\supset\! \Tilde{\X}_i \},
\mathsf{W}_i \!=\! \{\Tilde{w}_{ip} | \!\bigcup_{p=1}^{P}\! B_{\varepsilon_{iu}}(\Tilde{w}_{ip}) \!\supset\! \Tilde{\W}_i \}.
\end{align}
We introduce the following assumptions to formulate the main theorem of this subsection.
\begin{assumption}\label{assum:Lipschitz}
    For all $i \in \mathcal{I}_N$, the {conditions in \eqref{eq:l-ISS-LF} are Lipschitz continuous in $(\Tilde{x}_i, \hat{\Tilde{x}}_i, \Tilde{w}_i, \hat{\Tilde{w}}_i) \in \Tilde{\X}_i \times \Tilde{\X}_i \times \Tilde{\W}_i \times \Tilde{\W}_i$ with Lipschitz constants $\mathsf{L}_{ik}$ with $k \in [1;3]$ for the three conditions.} The maximum of these constants is $\mathsf{L}_i$.
\end{assumption}
\begin{remark}
    The Lipschitz continuity of these conditions can be enforced via the Lipschitz continuity of the subsystem dynamics \eqref{eq:subsystem_dyn_2} and the GNN $V_i^{\gamma}$ corresponding to the local $\delta$-ISS Lyapunov function. {Lipschitz constant of the subsystem dynamics are assumed to be known. Note that the GNN architecture defined in Section~\ref{subsec:GNN_archi} is inherently Lipschitz continuous due to linear operators and ReLU activation functions. A low Lipschitz constant may be enforced via $L2$ regularization or bounding of the spectral norm of the neural network weights. The spectral norm of the weight matrices are then utilized to compute the Lipschitz constant of the local $\delta$-ISS Lyapunov function.} 
\end{remark}
Now we state the main theorem of this subsection that provides a formal guarantee of the trained network over the entire state-space.
\begin{theorem}\label{th:guarantee}
    Consider a large-scale interconnected system \eqref{eq:complete} composed of $N$ homogeneous subsystems described by \eqref{eq:subsystem_dyn_2} with a topology according to $\mathcal{G}$ for a given state-space $\X$ and input-space $\W$. For each subsystem $i \in \mathcal{I}_N$, let $V_i^{\gamma}$ be the function representing the candidate local $\delta$-ISS Lyapunov function trained over the randomly sampled points collected from $\Tilde{\X}_i$ as discussed in Section \ref{subsec:training}. Then, the GNN $V_i^{\gamma}$ is a valid local $\delta$-ISS Lyapunov function satisfying condition \eqref{eq:l-ISS-LF} for the subsystem $i$ for all $i \in \mathcal{I}_N$ if the following conditions hold with $\hat{\eta}_i + \sqrt{2}\mathsf{L}_i \varepsilon_i \leq 0$ where $\varepsilon_i = \max(\varepsilon_{ix}, \varepsilon_{iu})$:
    \begin{subequations}\label{eq:SOP}
    \begin{align}
    & \forall (\Tilde{x}_{ir}, \hat{\Tilde{x}}_{ir}) \in \mathcal{X}_i \times \mathcal{X}_i, \Tilde{x}_{ir} \neq \hat{\Tilde{x}}_{ir}, \forall (\Tilde{w}_{ip}, \hat{\Tilde{w}}_{ip}) \in \mathsf{W}_i \times \mathsf{W}_i: \notag \\
        & \quad -V_i^{\gamma}(\Tilde{x}_{ir}, \hat{\Tilde{x}}_{ir}) + \underline \alpha_i |\Tilde{x}_{ir} - \hat{\Tilde{x}}_{ir}|^\kappa \leq \hat{\eta}_i, \label{eq:geq_SOP}\\
        & \quad V_i^{\gamma}(\Tilde{x}_{ir}, \hat{\Tilde{x}}_{ir}) - \overline \alpha_i |\Tilde{x}_{ir} - \hat{\Tilde{x}}_{ir}|^\kappa \leq \hat{\eta}_i, \label{eq:leq_SOP}\\
        & \quad V_i^{\gamma}(\Tilde{f}_i(\Tilde{x}_{ir}, \Tilde{w}_{ip}), \Tilde{f}_i(\hat{\Tilde{x}}_{ir}, \hat{\Tilde{w}}_{ip})) - V_i(\Tilde{x}_{ir}, \hat{\Tilde{x}}_{ir}) +\Tilde{\alpha}_{ir}|\Tilde{x}_{ir} - \hat{\Tilde{x}}_{ir}|^\kappa - \sigma_i|\Tilde{w}_{ip}-\hat{\Tilde{w}}_{ip}|^\kappa \leq \hat{\eta}_i \label{eq:diff_SOP}.
    \end{align}
    \end{subequations}
\end{theorem}
\begin{proof}
    For $i \in \mathcal{I}_N$, it is evident that with $\hat{\eta}_i$, the condition \eqref{eq:geq_SOP} has been satisfied for the points in $\mathcal{X}_i$. Now from \eqref{set:SCP}, it is evident there exists a $\Tilde{x}_{ir} \in \mathcal{X}_i$ for every $\Tilde{x}_i \in \Tilde{\X}_i$ with $|\Tilde{x}_i - \Tilde{x}_{ir}| \leq \varepsilon_i$. Therefore, under Assumption \ref{assum:Lipschitz}, it is straightforward to derive that for all $\Tilde{x}_i, \hat{\Tilde{x}}_i \in \Tilde{\X}_i, -V_i^{\gamma}(\Tilde{x}_i, \hat{\Tilde{x}}_i) + \underline \alpha_i |\Tilde{x}_i - \hat{\Tilde{x}}_i|^\kappa \leq -V_i^{\gamma}(\Tilde{x}_{i}, \hat{\Tilde{x}}_i) + \underline \alpha_i |\Tilde{x}_i - \hat{\Tilde{x}}_i|^\kappa -V_i^{\gamma}(\Tilde{x}_{ir}, \hat{\Tilde{x}}_{ir}) + \underline \alpha_i |\Tilde{x}_{ir} - \hat{\Tilde{x}}_{ir}|^\kappa + V_i^{\gamma}(\Tilde{x}_{ir}, \hat{\Tilde{x}}_{ir}) - \underline \alpha_i |\Tilde{x}_{ir} - \hat{\Tilde{x}}_{ir}|^\kappa \leq \hat{\eta}_i + \mathcal{L}_i|(\Tilde{x}_i, \hat{\Tilde{x}}_i) - (\Tilde{x}_{ir}, \hat{\Tilde{x}}_{ir})| \leq \hat{\eta}_i + \sqrt{2}\mathcal{L}_i\varepsilon_i \leq 0$, implying the satisfaction of first condition of \eqref{eq:l-ISS-LF}. The satisfaction of other conditions of \eqref{eq:l-ISS-LF} follows a similar approach, which completes the proof.
\end{proof}

{\textbf{Computation complexity:} The computational complexity of verifying distributed Lyapunov functions ensuring $\delta$-ISS is primarily determined by the number of data samples collected from the augmented space, which involves quantization of state and input space with respect to the neighborhood of every subsystem only. As described in \eqref{set:SCP}, the augmented state-space has a dimension of $n + n_{\mathcal{N}_i}$ and the input-space has a dimension of $m + m_{\mathcal{N}_i}$. Considering the maximum number of neighbours for any subsystem is $d$, \ie, $\max_{i=1}^N \mathcal{N}_i \leq d$, therefore the verification dataset size will be $R^{2n(d+1)}\times P^{2m(d+1)}$ where $R,P$ are the number of samples per dimension collected from the state-space and input-space, respectively.}
\begin{remark}
    It should be carefully noted that the number of samples required for verifying distributed $\delta$-ISS Lyapunov functions is significantly lower than the centralized method \cite{basu2025formally}. In centralized certificate verification, {one must consider quantization for the entire large-scale system, thereby requiring a total of $R^{2Nn}\times P^{2Nm}$ samples for verification, which clearly shows that the size of the verification dataset is significantly larger compared to the proposed verification technique. Therefore, it is evident that} the scalability of this verification technique depends on the sparsity of the adjacency matrix $\mathsf{A}$. For a fully interconnected structure, our approach of verification would still require the same number of samples as the centralized one. To deal with such issues, one may be able to utilize an appropriate learning framework for compositional construction of $\delta$-ISS Lyapunov functions under some \textit{small-gain} assumptions on the interconnections \cite{DD-Stability}, which can be a good direction of future work.
\end{remark}
\begin{remark}
    Note that, $\hat{\eta}_i$ in \eqref{eq:SOP} and $\lambda_i$ in \eqref{eq:loss} serve a similar purpose in extending the conditions on the Lyapunov function beyond the finite dataset for verification. In practice, we directly examine the satisfaction of the condition $\hat{\eta}_i + \sqrt{2}\mathsf{L}_i\varepsilon_i \leq 0$ inside the training procedure with $\hat{\eta}_i = \lambda_i$ which, when satisfied, implies the trained GNN is valid for the entire state-space.
\end{remark}

\section{Case Study}
We validate the effectiveness of the proposed approach over two case studies: the first being a temperature control network, while the second one is a network of 2-D nonlinear system. All the computations were performed using PyTorch in Python 3.10 by leveraging GNN architectures from \cite{Gama2019GNN} on a Windows-operated machine with an Intel Core i7-14700 CPU and 32 GB RAM.

\begin{figure}[h]
    \centering
    \includegraphics[width=\linewidth]{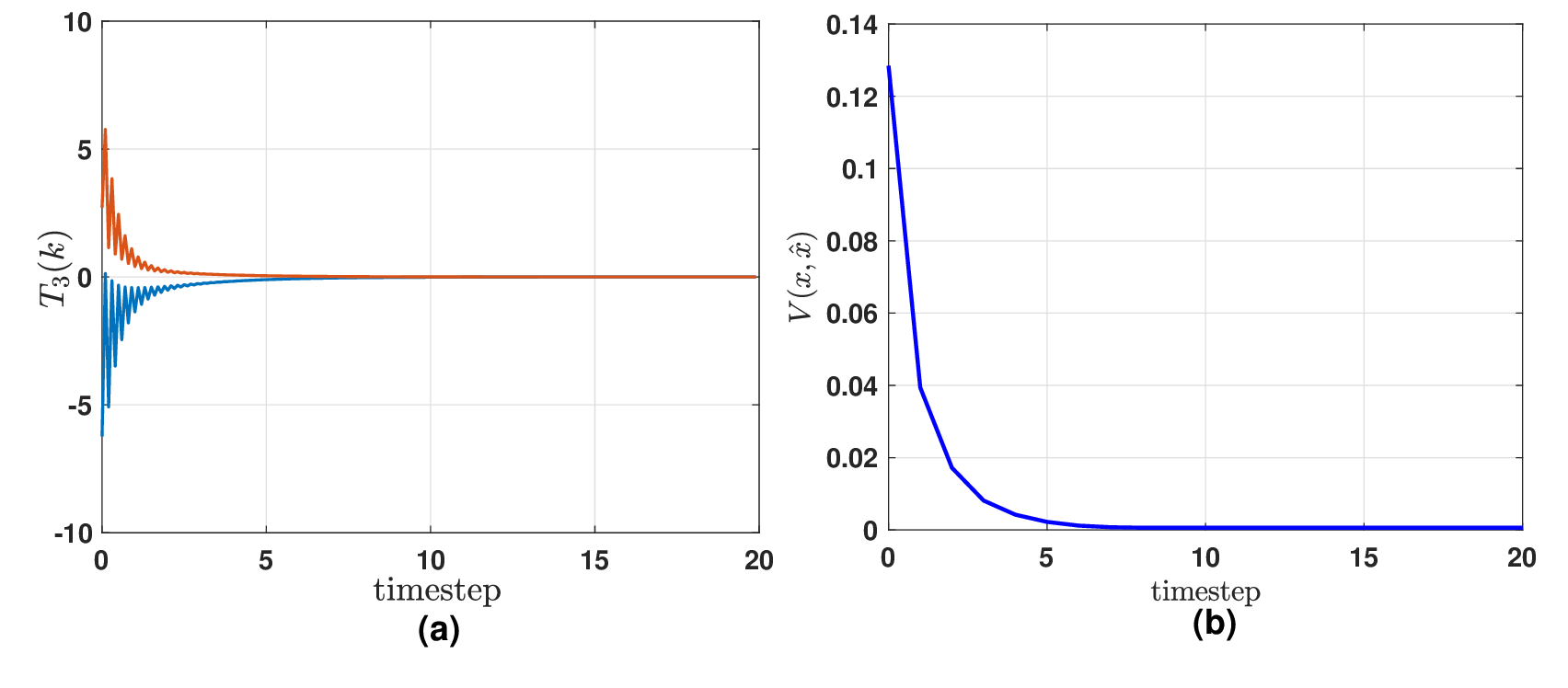}
    \vspace{-0.8cm}
    \caption{{(a) Trajectories of a subsystem starting from different initial conditions converge towards each other. (b) The Lyapunov value over the trajectory decays, indicating $\delta$-GAS in the absence of external input.}}
    \label{fig:sim1}
\end{figure}

\textbf{Temperature Control Network:} For the first case study, we consider a temperature control network comprising $N$ rooms in a circular topology where each room temperature $T_i(\cdot), i \in \mathcal{I}_N$ evolution can be described by \cite{meyer2017compositional}:
\begin{align*}
    \Upsilon: T_i(k+1) &= T_i(k) + \phi(T_{i+1}(k) +T_{i-1}(k) - 2T_i(k)) + \theta(T_e - T_i(k)), 
\end{align*}
where $\phi, \theta$ are the thermal factors between neighbouring rooms and external environment, respectively and $T_e$ is the external temperature. Therefore, we consider the state-space of the large-scale system to be $\X = [-10, 10]^N$ while the model is unknown. The interconnection topology $\mathcal{G}$ is described by the adjacency matrix $\mathsf{A}$ such that for all $i \in \mathcal{I}_N, [\mathsf{A}]_{ii} = [\mathsf{A}]_{1N} = [\mathsf{A}]_{N1} = [\mathsf{A}]_{i,i+1} = [\mathsf{A}]_{i+1,i}=1$ and all other elements are zero. For the GNN corresponding to the distributed $\delta$-ISS Lyapunov function, we consider $\bar{l}_b=1, l_b = 2$ and $ h_b^j=20$ for $i \in [1;2]$. We train the GNN to obtain the local $\delta$-ISS Lyapunov function $V_i^\gamma$ satisfying condition \eqref{eq:l-ISS-LF} for $\eta_i = -0.0003, \ i \in \mathcal{I}_N$. We first train the certificates for $N=10$ and then consequently, fine-tune the trained Lyapunov function for $N=50,100,500,1000$. This is possible due to repeated application of training parameters at every node of the graph as well as its regularity. The maximum of the Lipschitz constant of the conditions is obtained as $\mathcal{L}_i = 1.25$. Finally, we verify the correctness of the functions by considering $\varepsilon_i = 0.0002$. Thus, we obtain the $\delta$-ISS Lyapunov function for the interconnected system $V = \sum_{i=1}^N V_i^\gamma$. Figure \ref{fig:sim1}(a) shows the {trajectory of a subsystem converging towards each other while \ref{fig:sim1}(b) shows the decay of Lyapunov value over the trajectory in absence of external input, thereby verifying the interconnected system to be $\delta$-GAS.}
The time taken for training with verification are 45 minutes for $N=10$ and an additional 20 minutes for fine-tuning up to $N=1000$. 

\begin{figure}[hbt!]
    \centering
    \includegraphics[width=\linewidth]{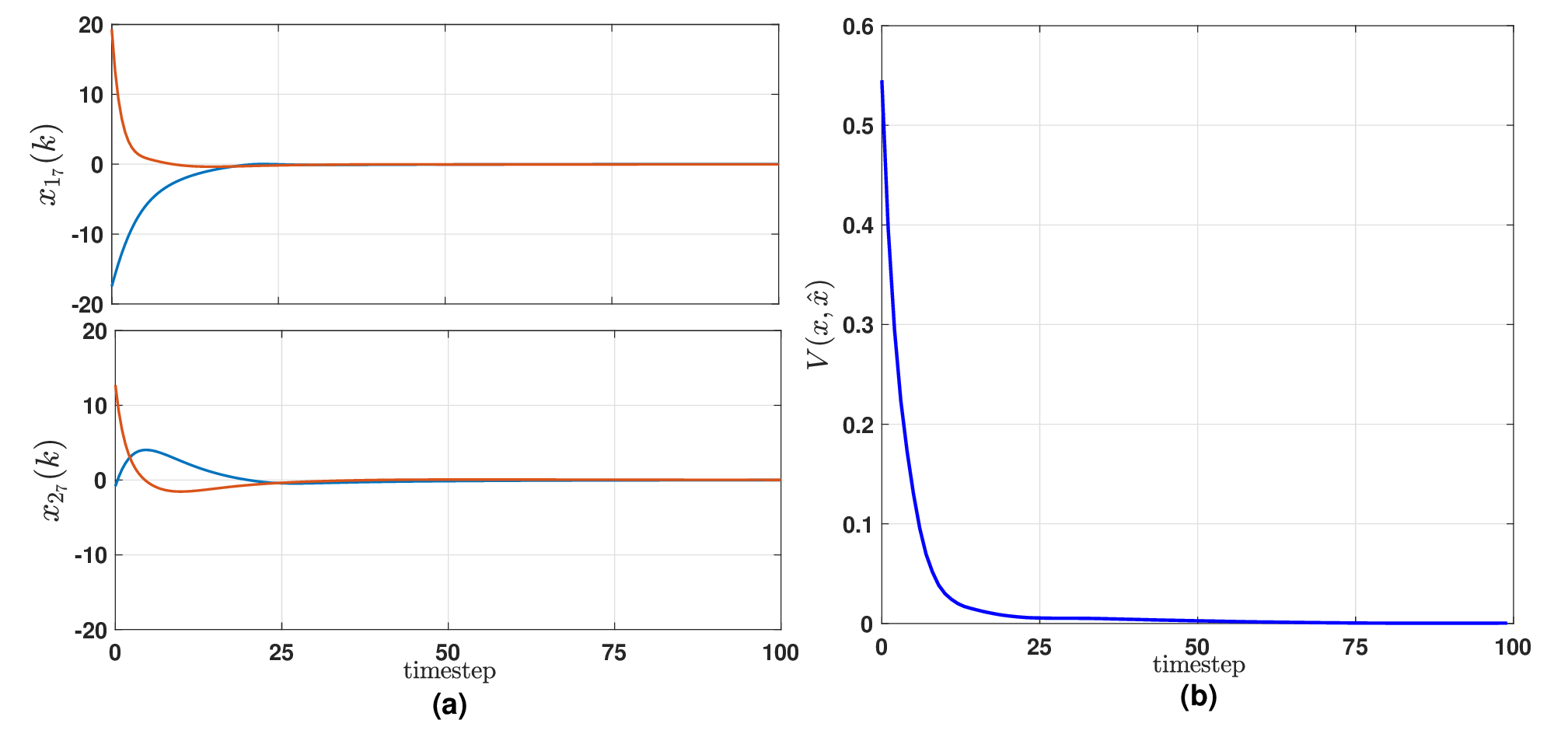}
    \vspace{-0.75cm}
    \caption{{(a) Trajectories of a subsystem starting from different initial conditions converge towards each other. (b) The Lyapunov value over the trajectory decays, indicating $\delta$-GAS in the absence of external input.}}
    \label{fig:sim2}
\end{figure}

\textbf{Nonlinear 2D system:} We consider a nonlinear two-dimensional system where each subsystem possesses the following dynamics:
\begin{align*}
    x_{1_i}(k+1) &= 0.8x_{1_i}(k) - 0.1 \sqrt{x_{1_i}^2(k) + x_{2_i}^2(k)} - 0.02 x_{1_{i+1}}(k) \\
    x_{2_i}(k+1) &= 0.9x_{2_i}(k) - 0.1 x_{1_i}(k) - 0.03x_{2_{i+1}}(k),
\end{align*}
in which it is evident that the subsystems are interconnected through a ring topology but in one direction only, unlike the previous case study. Therefore, the interconnection topology $\mathcal{G}$ is described by the adjacency matrix $\mathsf{A}$ such that for all $i \in \mathcal{I}_N, [\mathsf{A}]_{ii} =[\mathsf{A}]_{N1} = [\mathsf{A}]_{i,i+1} = 1$ and all other elements are zero. We consider the state-space of the large-scale system to be $\X = [-20, 20]^N$ while the model is unknown. For the GNN corresponding to the distributed $\delta$-ISS Lyapunov function, we consider $\bar{l}_b=1, l_b = 2$ and $ h_b^j=20$ for $i \in [1;2]$.  The maximum of the Lipschitz constant of the conditions is obtained as $\mathcal{L}_i = 1.5$ while we consider $\varepsilon_i = 0.001$ for verifying the correctness of the trained Lyapunov function.
Figure \ref{fig:sim2}(a) shows the {trajectories of a subsystem converging towards each other, while \ref{fig:sim2}(b) shows the decay of Lyapunov value over the trajectory satisfying the incremental stability property of the interconnected system.} The time taken for training with verification is 2 hours for $N=10$ and an additional 40 minutes for fine-tuning up to $N=1000$. 

\section{Conclusion and Future Work}
This work presents a framework for constructing formally verified neural $\delta$-ISS Lyapunov functions for large-scale interconnected systems through a distributed, graph-based representation. By leveraging the notion of local $\delta$-ISS Lyapunov function, we introduce a compositionality condition for constructing a Lyapunov function for the overall interconnected system. Then we use GNN architecture to approximate the local Lyapunov functions and propose a data-driven verification scheme to formally verify the trained certificates over the complete state-space. A possible future direction of the work is to synthesize GNN-based local controllers ensuring incremental stability of large-scale interconnected system that unifies stability certification and control design within a single learning-based, distributed framework.

\bibliographystyle{plain}
\bibliography{sources}

\end{document}